\documentclass[pra,aps,eqsecnum,amsmath,amssymb,showkeys,showpacs]{revtex4}
\usepackage[mathscr]{eucal}
\usepackage{xtheorem}
\usepackage{graphicx}

\newcommand{\pbi}{\pmb{\mathbb{I}}}
\newcommand{\pp}{\mathsf{p}}
\newcommand{\qp}{\mathsf{q}}
\newcommand{\cp}{\mathsf{c}}
\newcommand{\pim}{\mathsf{\Pi}}

\newcommand{\am}{\mathsf{A}}
\newcommand{\bn}{\mathsf{B}}
\newcommand{\xm}{\mathsf{X}}
\newcommand{\ym}{\mathsf{Y}}
\newcommand{\mm}{\mathsf{M}}
\newcommand{\nm}{\mathsf{N}}
\newcommand{\um}{\mathsf{U}}
\newcommand{\vm}{\mathsf{V}}
\newcommand{\dm}{\mathsf{D}}
\newcommand{\sm}{\mathsf{S}}
\newcommand{\jm}{\mathsf{J}}
\newcommand{\ax}{\mathsf{X}}
\newcommand{\bsi}{\boldsymbol{\psi}}
\newcommand{\bhi}{\boldsymbol{\varphi}}
\newcommand{\tr}{{\rm{tr}}}
\newcommand{\ac}{\mathscr{A}}
\newcommand{\bc}{\mathscr{B}}
\newcommand{\mc}{\mathcal{M}}
\newcommand{\nc}{\mathcal{N}}
\newcommand{\hh}{\mathcal{H}}
\newcommand{\kk}{\mathcal{K}}
\newcommand{\ri}{\mathrm{i}}
\newtheorem{plain}{Thm}{Theorem}[section]
{Lemma}
{Example}
\begin{document}
\clearpage
\preprint{}

\title{Entropic uncertainty relations for extremal unravelings of super-operators}
\author{Alexey E. Rastegin}
\affiliation{Department of Theoretical Physics, Irkutsk State University,
Gagarin Bv. 20, Irkutsk 664003, Russia}

\begin{abstract}
A way to pose the entropic uncertainty principle for
trace-preserving super-operators is presented. It is based on the
notion of extremal unraveling of a super-operator. For given input
state, different effects of each unraveling result in some
probability distribution at the output. As it is shown, all
Tsallis' entropies of positive order as well as some of
R\'{e}nyi's entropies of this distribution are minimized by the
same unraveling of a super-operator. Entropic relations between a
state ensemble and the generated density matrix are revisited in
terms of both the adopted measures. Using Riesz's theorem, we
obtain two uncertainty relations for any pair of generalized
resolutions of the identity in terms of the R\'{e}nyi and Tsallis
entropies. The inequality with R\'{e}nyi's entropies is an
improvement of the previous one, whereas the inequality with
Tsallis' entropies is a new relation of a general form. The latter
formulation is explicitly shown for a pair of complementary
observables in a $d$-level system and for the angle and the
angular momentum. The derived general relations are immediately
applied to extremal unravelings of two super-operators.
\end{abstract}

\maketitle

\pagenumbering{arabic}
\setcounter{page}{1}

\section{Introduction}

Since the famous Heisenberg's paper \cite{heisenberg} had been
published, uncertainty relations are the subject of long
researches \cite{hall99,lahti}. A new interest was stimulated by
recent advances in quantum information processing. There are two
well-known approaches to formulating the uncertainty principle.
The first was initiated by Robertson \cite{robert} who showed that
a product of the standard deviations of two observables is bounded
from below. Here we may run across some disputable topics such as
the number-phase case \cite{lynch}. Both the well-defined
Hermitian operator of phase and corresponding number-phase
uncertainty relation have been fit within the Pegg-Barnett
formalism \cite{barnett}. The second approach is generally
characterized by posing the uncertainty principle via
information-theoretic terms especially via entropies
\cite{maass,ww10}. Although the first relation of such a kind was
derived by Hirschman \cite{hirs}, a general statement of the
problem is examined in the papers \cite{mamojka,deutsch}. Mutually
unbiased bases \cite{ivan92,sanchez93}, the time-energy case
\cite{hall08} and tomographic processes \cite{manko09} have been
considered within an entropic approach as well.

The most of known uncertainty relations deals with observables or,
more generally, with POVM measurements. Nevertheless, there exist
relations for unitary transformations \cite{spindel} and non-Hermitian
annihilation operator \cite{lanz}. Both the measurement and
unitary evolution are rather simplest types of a state change in
quantum theory \cite{kraus}. The formalism of quantum operations,
or super-operators, is now a standard tool for treating quantum
processes. In the present work, we address a question how to
formulate the uncertainty principle for super-operators. It turns
out that one of possible ways is naturally provided with the
notion of extremal unraveling of a super-operator. For the Shannon
entropy, this notion was examined in \cite{ilichev03}. As an
entropic measure, we will use the Tsallis entropy, which has found
use in various physical problems (see references in
\cite{tsbibl}), and the R\'{e}nyi entropy.

The paper is organized as follows. In Section \ref{sec2}, the
properties of the Tsallis entropies and super-operators are
recalled. For given input state and super-operator, we find the
extremal unraveling that minimizes all the Tsallis entropies
simultaneously. Relations between the ensemble entropy and the
entropy of a generated density matrix are revisited. In Section
\ref{sec3}, we derive the uncertainty relation for two generalized
resolutions of the identity in terms of Tsallis entropies. The
previous result on the R\'{e}nyi entropies is refined. The obtained
entropic relations are directly used for the extremal unravelings of
two super-operator. Section \ref{sec4} concludes the paper with a summary of results.

\section{Tsallis' entropies and extremal unravelings}\label{sec2}

First, we briefly recall the definitions of used entropic measures
(for a discussion and further references, see \cite{aczel}). For
real $\alpha>0$ and $\alpha\neq1$, we define the non-extensive
$\alpha$-entropy of probability distribution $\{p_i\}$ by
\cite{tsallis}
\begin{equation}
H_{\alpha}(p_i)\triangleq(1-\alpha)^{-1} \left(\sum\nolimits_i p_i^{\alpha}
- 1 \right) \ . \label{tsaent}
\end{equation}
This can be rewritten as $H_{\alpha}(p_i)=-\sum_i
p_i^{\alpha}\ln_{\alpha}p_i$, where
$\ln_{\alpha}x\equiv\bigl(x^{1-\alpha}-1\bigr)/(1-\alpha)$ is the
$\alpha$-logarithmic function, defined for $\alpha\geq0$,
$\alpha\neq1$ and $x\geq0$. The quantity (\ref{tsaent})
will be referred to as ''Tsallis $\alpha$-entropy'', though it was
previously discussed by Havrda and Charv\'{a}t \cite{havrda}. In the
limit $\alpha\to1$, $\ln_{\alpha}x\to\ln{x}$ and the quantity
(\ref{tsaent}) recovers the Shannon entropy. We will also use the
R\'{e}nyi $\alpha$-entropy defined for $\alpha\neq1$ as
\cite{renyi}
\begin{equation}
R_{\alpha}(p_i)\triangleq(1-\alpha)^{-1} \ln\left(\sum\nolimits_i p_i^{\alpha}
\right) \ . \label{renent}
\end{equation}
The R\'{e}nyi $\alpha$-entropy also coincides with the Shannon
entropy in the limit $\alpha\to1$. There are many various forms of
extrapolation between different entropies \cite{zycz}. We will
only need in the equality
\begin{equation}
R_{\alpha}(p_i)=(1-\alpha)^{-1} \ln\Bigl( 1+(1-\alpha)H_{\alpha}(p_i)
\Bigr) \ , \label{rentsa}
\end{equation}
which immediately follows from the definitions (\ref{tsaent}) and
(\ref{renent}).

In the following, some notation of linear algebra will be used.
Let $\hh$ be finite-dimensional Hilbert space. For given two
vectors $\bsi,\bhi\in\hh$, their inner product is denoted by
$\langle\bsi{\,},\bhi\rangle$. For two operators $\xm$ and $\ym$
on $\hh$, we define the Hilbert--Schmidt inner product by
\cite{watrous1}
\begin{equation}
\langle\xm{\,},\ym\rangle_{\rm{hs}}
\triangleq{\rm{tr}}(\xm^{\dagger}\ym)
\ . \label{hsdef}
\end{equation}
This inner product naturally induces so-called Frobenius norm
$\|\xm\|_F=\langle\xm{\,},\xm\rangle_{\rm{hs}}^{1/2}$. The
Frobenius norm can be re-expressed as
$\|\xm\|_F=\Bigl(\sum_js_j(\xm)^2\Bigr)^{1/2}$, where the singular
values $s_j(\xm)$ are eigenvalues of
$|\xm|=\sqrt{\xm^{\dagger}\xm}$. The largest singular value of
$\xm$ gives the spectral norm $\|\xm\|_{\infty}$ of this operator
\cite{watrous1}.

A quantum measurement is described by a ''generalized resolution
of the identity'' (or by a ''positive operator-valued measure'').
This is a set $\{{\mathsf{M}}_i\}$ of positive semidefinite
operators obeying the completeness relation $\sum_i
{\mathsf{M}}_i=\pbi$, where $\pbi$ is the identity \cite{holevo}.
Consider a linear map $\$$ that takes linear operators on $\hh$ to
linear operators on $\hh'$, and  also satisfies the conditions of
trace preservation and complete positivity. Following
\cite{preskill}, this map will be referred to as
''super-operator''. Each super-operator has a Kraus
representation, namely \cite{kraus,preskill}
\begin{equation}
\$(\rho)=\sum\nolimits_i \am_i{\,}\rho{\,}\am_i^{\dagger}
\ , \label{opsum}
\end{equation}
where the Kraus operators $\am_i$ map the input space $\hh$ to the
output space $\hh'$ and obey $\sum_i\am_i^{\dagger}\am_i=\pbi$
(the preservation of the trace). Representations of such a kind
are never unique \cite{watrous1}. For given super-operator $\$$,
there are many sets $\ac=\{\am_i\}$ that enjoy (\ref{opsum}). In
the paper \cite{ilichev03}, each concrete set $\ac=\{\am_i\}$
resulting in (\ref{opsum}) is called an ''unraveling'' of the
super-operator $\$$. This terminology is due to Carmichael
\cite{carm} who introduced this word for a representation of the
master equation. It is well-known that two Kraus representations
of the same super-operator are related as
\begin{equation}
\bn_i=\sum\nolimits_j \am_j{\,}u_{ji}
\ , \label{eqvun}
\end{equation}
where the matrix $\um=[[u_{ij}]]$ is unitary \cite{preskill}. [We
assume in (\ref{eqvun}) that unravelings $\ac$ and $\bc$ have the
same cardinality by adding zero operators, if needed.] For given
density operator $\rho$ and super-operator unraveling
$\ac=\{\am_j\}$, we introduce the matrix
\begin{equation}
\pim(\ac|\rho)\triangleq[[\langle\am_i\sqrt{\rho}{\,},\am_j\sqrt{\rho}\rangle_{\rm{hs}}]]\equiv
[[\tr(\am_i^{\dagger}\am_j\rho)]]
\ . \label{pimdef}
\end{equation}
The diagonal element $p_i=\tr(\am_i^{\dagger}\am_i\rho)$ is
clearly positive and gives the $i$th effect probability. Then the
entropies $H_{\alpha}(\ac|\rho)$ and $R_{\alpha}(\ac|\rho)$ are
merely defined by (\ref{tsaent}) and (\ref{renent}) respectively.
By definition, the matrix $\pim(\ac|\rho)$ is Hermitian. Suppose
that the two sets $\ac=\{\am_i\}$ and $\bc=\{\bn_j\}$ fulfill
(\ref{eqvun}). Using the properties of the inner product, we have
\begin{equation}
\langle\bn_i\sqrt{\rho}{\,},\bn_k\sqrt{\rho}\rangle_{\rm{hs}}=
\sum\nolimits_{jl} u_{ji}^{*}{\,}u_{lk}{\,}
\langle\am_j\sqrt{\rho}{\,},\am_l\sqrt{\rho}\rangle_{\rm{hs}}
\ , \label{pinner}
\end{equation}
or $\pim(\bc|\rho)=\um^{\dagger}{\,}\pim(\ac|\rho){\,}\um$ as the
matrix relation. That is, if the sets $\ac$ and $\bc$ are both
unravelings of the same super-operator then the matrices
$\pim(\ac|\rho)$ and $\pim(\bc|\rho)$ are unitarily similar. Due
to Hermiticity, all the matrices of a kind $\pim(\ac|\rho)$
assigned to one and the same super-operator are unitarily similar
to a unique (up to permutations) diagonal matrix
$\dm=\rm{diag}(\lambda_1,\lambda_2,\ldots)$, where the
$\lambda_i$'s and perhaps zeros are the eigenvalues of each of
these matrices. So any $\pim(\ac|\rho)$ is positive semidefinite.

For given unraveling $\ac=\{\am_i\}$, we obtain the concrete
matrix $\pim(\ac|\rho)$ and diagonalize it through a unitary
transformation $\vm^{\dagger}{\,}\pim(\ac|\rho){\,}\vm=\dm$. 
Let us define a specific unraveling $\ac_{\rho}^{(ex)}$ related to
given $\ac$ as
\begin{equation}
\am_{i}^{(ex)}=\sum\nolimits_j \am_j{\,} v_{ji}
\ , \label{excalc}
\end{equation}
where the unitary matrix $\vm$ diagonalizes $\pim(\ac|\rho)$. It turns 
out that the unraveling (\ref{excalc}) enjoys the {\it extremality} property with respect to all the
Tsallis $\alpha$-entropies for $\alpha\in(0;\infty)$ and the
R\'{e}nyi $\alpha$-entropies for $\alpha\in(0;1)$. 

\begin{Thm}\label{extr}
For given density operator $\rho$ and super-operator $\$$, each
unraveling $\ac$ of the $\$$ satisfies
\begin{eqnarray}
 & H_{\alpha}(\ac_{\rho}^{(ex)}|\rho)\leq H_{\alpha}(\ac|\rho) {\quad} \forall{\ } \alpha\in(0;\infty)
\ , \label{tsaex} \\
 & R_{\alpha}(\ac_{\rho}^{(ex)}|\rho)\leq R_{\alpha}(\ac|\rho) {\quad} \forall{\ } \alpha\in(0;1)
\ , \label{renex}
\end{eqnarray}
where the extremal unraveling $\ac_{\rho}^{(ex)}$ is defined by
the formula (\ref{excalc}).
\end{Thm}

{\bf Proof} We firstly note that
$\pim(\ac_{\rho}^{(ex)}|\rho)=\dm$ with the probabilities
$\lambda_j$ of effects. In view of
$\pim(\ac|\rho)=\vm{\,}\dm{\,}\vm^{\dagger}$, the probabilities
$p_i=\tr(\am_i^{\dagger}\am_i\rho)$ of different effects of the
unraveling $\ac$ is related to the $\lambda_j$'s by
\begin{equation}
p_i=\sum\nolimits_j v_{ij}{\,}\lambda_j{\,}v_{ij}^{*} =\sum\nolimits_j s_{ij}{\,}\lambda_j
\ , \label{pipij}
\end{equation}
where $s_{ij}=v_{ij}{\,}v_{ij}^{*}$. The matrix $\sm=[[s_{ij}]]$
is unistochastic, whence $\sum\nolimits_{i}s_{ij}=1$ for all
$j$ and $\sum\nolimits_{j}s_{ij}=1$ for all $i$. We now use
these relations and an obvious fact that for $\alpha>0$ the
function $h_{\alpha}(x)=\bigl(x^{\alpha}-x\bigr)\big/(1-\alpha)$
is concave in the range $x\in[0;1]$. According to Jensen's
inequality (see, e.g., the book \cite{hardy}), there holds
\begin{equation}
H_{\alpha}(\ac|\rho)=\sum\nolimits_i h_{\alpha}\Bigl(\sum\nolimits_js_{ij}{\,}\lambda_j\Bigr)\geq
\sum\nolimits_i \sum\nolimits_js_{ij}{\,}h_{\alpha}(\lambda_j)=
\sum\nolimits_jh_{\alpha}(\lambda_j)=H_{\alpha}(\ac_{\rho}^{(ex)}|\rho)
\ . \label{jens}
\end{equation}
This completes the proof for (\ref{tsaex}). Further, we note that
for $\alpha<1$ the function
$(1-\alpha)^{-1}\ln\bigl(1+(1-\alpha){\,}x\bigr)$ is increasing in
the range $x\in[0;\infty)$. Combining the equality (\ref{rentsa})
with (\ref{tsaex}) then gives (\ref{renex}). $\square$

For the Shannon entropy, a problem of ''minimal'' unraveling
was considered in \cite{breslin} and later in
\cite{ilichev03}. Diagonalizing the matrix $\pim(\ac|\rho)$ is
actually equivalent to the extreme condition that has been derived by
the method of Lagrange's multipliers in \cite{ilichev03}. Latter
reasons local in spirit are quite complemented by the above proof
based on the concavity. Our treatment allows to find the extremal
unraveling easily from (\ref{excalc}). Thus, for prescribed state
$\rho$ all the Tsallis entropies of order $\alpha\in(0;\infty)$
and the R\'{e}nyi entropies of order $\alpha\in(0;1)$ are
minimized by one and the same unraveling of given super-operator.
However, this unraveling does not minimize other  R\'{e}nyi
entropies in general. An unraveling extremal with respect to
R\'{e}nyi's entropy of order $\alpha>1$ may also be dependent on
$\alpha$. A search for such an unraveling seems to
be difficult since R\'{e}nyi's entropy of such an order is not
purely convex nor purely concave.

The relations (\ref{tsaex}) and (\ref{renex}) can be put in the
context of a state ensemble having a prescribed density operator.
In line with (\ref{tsaent}) and (\ref{renent}), we
introduce the quantum Tsallis and R\'{e}nyi entropies of a density
matrix $\rho$ by
\begin{equation}
{\rm{H}}_{\alpha}(\rho)\triangleq(1-\alpha)^{-1}\tr\bigl(\rho^{\alpha}-\rho\bigr)
\ , {\ }\qquad {\rm{R}}_{\alpha}(\rho)\triangleq(1-\alpha)^{-1}\ln\bigl(\tr(\rho^{\alpha})\bigr)
\ . \label{qentdf}
\end{equation}
In the limit $\alpha\to1$ both the expressions coincides with the
von Neumann entropy $-\tr\bigl(\rho\ln\rho\bigr)$. Let
$\{p_i,\bsi_i\}$ be an ensemble of normalized pure states, $\langle\bsi_i{\,},\bsi_i\rangle=1$ and
$\sum_ip_i=1$, that leads to the density operator $\rho$, namely
\begin{equation}
\sum\nolimits_i p_i{\,}\bsi_i{\,}\bsi_i^{\dagger}=\rho=
\sum\nolimits_j \lambda_j{\,}\bhi_j{\,}\bhi_j^{\dagger}
\ . \label{ensres}
\end{equation}
In general, the states $\bsi_i$ are not mutually orthogonal. The
right-hand side of (\ref{ensres}) poses the spectral decomposition of
$\rho$, so that the vectors $\bhi_j$ form an orthonormal basis in
$\hh$. The ensemble classification theorem says that
\cite{hugston}
\begin{equation}
\sqrt{p_i}{\,}\bsi_i=\sum\nolimits_j u_{ij}\sqrt{\lambda_j}{\,}\bhi_j
\label{unfr}
\end{equation}
for some unitary matrix $[[u_{ij}]]$. It follows from
(\ref{unfr}) and $\langle\bhi_j{\,},\bhi_k\rangle=\delta_{jk}$
that $p_i=\sum_js_{ij}{\,}\lambda_j$, where $s_{ij}=u_{ij}^{*}u_{ij}$ 
are elements of a unistochastic matrix. Changing notation in the formula
(\ref{jens}) appropriately, we finally get
\begin{equation}
{\rm{H}}_{\alpha}(\rho)=H_{\alpha}(\lambda_j)\leq  H_{\alpha}(p_i) {\>}\quad (0<\alpha)\ ,
{\ }\qquad {\rm{R}}_{\alpha}(\rho)=R_{\alpha}(\lambda_j)\leq  R_{\alpha}(p_i) {\>}\quad (0<\alpha<1)
\ . \label{concpur}
\end{equation}
In the limit $\alpha\to1$, both the inequalities are reduced to
$-\tr\bigl(\rho\ln\rho\bigr)\leq H_1(p_i)$. The last is a known
relation between the von Neumann entropy of the generated density
operator and the Shannon entropy of an ensemble \cite{wehrl}. For
the Tsallis entropies this result can be proceeded to ensembles of
mixed states. Let an ensemble $\{p_i,\omega_i\}$ of normalized
density operators, $\tr(\omega_i)=1$ and
$\sum_ip_i=1$, give rise to a density
operator
\begin{equation}
\rho=\sum\nolimits_{i}p_i{\,}\omega_i=\sum\nolimits_{ij} p_i{\,}\nu_{ij}{\,}\bhi_{ij}{\,}\bhi_{ij}^{\dagger}
\ , \label{ensmix}
\end{equation}
where the spectral decomposition
$\omega_i=\sum_j\nu_{ij}{\,}\bhi_{ij}{\,}\bhi_{ij}^{\dagger}$.
Applying the first inequality of (\ref{concpur}) to the
right-hand side of (\ref{ensmix}), we have
$$
{\rm{H}}_{\alpha}(\rho)\leq-\sum_{ij}p_i^{\alpha}\nu_{ij}^{\alpha}\ln_{\alpha}(p_i\nu_{ij})
=-\sum_{ij}p_i^{\alpha}\nu_{ij}^{\alpha}\bigl(\ln_{\alpha}\nu_{ij}+\nu_{ij}^{1-\alpha}\ln_{\alpha}p_i\bigr)=
-\sum_i p_i^{\alpha} \sum_j \nu_{ij}^{\alpha}\ln_{\alpha}\nu_{ij}-\sum_i p_i^{\alpha}\ln_{\alpha}p_i \ .
$$
Here we used the identity $\ln_{\alpha}(xy)\equiv\ln_{\alpha}y+y^{1-\alpha}\ln_{\alpha}x$. Adding a consequence of concavity of the function
$h_{\alpha}(x)=\bigl(x^{\alpha}-x\bigr)\big/(1-\alpha)$, we finally write
\begin{equation}
\sum\nolimits_i p_i{\,}{\rm{H}}_{\alpha}(\omega_i)
\leq{\rm{H}}_{\alpha}(\rho)
\leq\sum\nolimits_i p_i^{\alpha}{\,}{\rm{H}}_{\alpha}(\omega_i)+H_{\alpha}(p_i)
\ . \label{concmix}
\end{equation}
The inequality on the left can be shown as follows. For any concave function $f(x)$, the functional
$\tr\bigl(f(\ax)\bigr)$ is also concave on Hermitian $\ax$ (for details, see section III in \cite{rast10fn}). We further note that ${\rm{H}}_{\alpha}(\rho)=\tr\bigl(h_{\alpha}(\rho)\bigr)$.
In the limit $\alpha\to1$, the inequalities (\ref{concmix}) are reduced to the
well-known bounds on the von Neumann entropy of a state mixture
(for instance, see \cite{wehrl}). It seems that such a treatment
fails in the case of R\'{e}nyi's entropies. Indeed, the Tsallis
$\alpha$-entropy does enjoy so-called strong additivity of degree
$\alpha$, whereas the R\'{e}nyi $\alpha$-entropy does not
\cite{aczel}.

\section{Entropic uncertainty relations}\label{sec3}

To obtain entropic relations, we will use a version of Riesz's
theorem (see theorem 297 in the book \cite{hardy}). Let
${\mathsf{x}}\in{\mathbb{C}}^n$ be $n$-tuple of complex numbers
$x_j$ and let ${\mathsf{T}}=[[t_{ij}]]$. Define $\eta$ to be
maximum of $|t_{ij}|$, i.e.
$\eta\triangleq{\max}\left\{|t_{ij}|:{\,}1\leq{i}\leq{m},1\leq{j}\leq{n}\right\}$.
The fixed matrix ${\mathsf{T}}$ describes a linear transformation
${\mathbb{C}}^n\rightarrow{\mathbb{C}}^m$. That is, to each
${\mathsf{x}}$ we assign $m$-tuple ${\mathsf{y}}\in{\mathbb{C}}^m$
with elements
\begin{equation}
y_i(x)=\sum\nolimits_{j=1}^{n} t_{ij}\> x_j \quad (i=1,\ldots,m)
\ . \label{lintrans}
\end{equation}
For $b\geq1$, the $l_b$ norm of vector ${\mathsf{x}}$ is
$\|{\mathsf{x}}\|_b=\left(\sum_j|x_j|^b\right)^{1/b}$. For $\beta\geq1/2$, we use a like function
$\|\qp\|_{\beta}=\left(\sum_jq_j^{\beta}\right)^{1/{\beta}}$ of probability distribution $\qp=\{q_j\}$,
though it is not a norm for $\beta<1$. Riesz's theorem is formulated as follows.

\begin{Lem}\label{riesz}
Suppose the matrix ${\mathsf{T}}$ satisfies
$\|{\mathsf{y}}\|_2\leq\|{\mathsf{x}}\|_2$ for all
${\mathsf{x}}\in{\mathbb{C}}^n$, $1/a+1/b=1$ and $1<b<2$; then
\begin{equation}
\|{\mathsf{y}}\|_{a}\leq{\eta}^{(2-b)/b}\|{\mathsf{x}}\|_{b}
\ . \label{suppos}
\end{equation}
\end{Lem}

We will now obtain an improved version of the statement emerged in
the paper \cite{rast102}. To each generalized resolution
$\mc=\{\mm_i\}$ and given mixed state $\rho$, we assign the
probabilistic vector $\pp=\{p_i\}$ with elements
$p_i=\tr(\mm_i\rho)\equiv\bigl\|\mm_i^{1/2}\sqrt{\rho}\bigr\|_F^{{\,}2}$.
Another resolution $\nc=\{\nm_j\}$ is assigned by the vector $\qp$
with elements
$q_j=\tr(\nm_j\rho)\equiv\bigl\|\nm_j^{1/2}\sqrt{\rho}\bigr\|_F^{{\,}2}$.

\begin{Lem}\label{resol}
For any two resolutions $\mc=\{\mm_i\}$ and $\nc=\{\nm_j\}$ of the
identity and given density operator $\rho$, there holds
\begin{equation}
\|\pp\|_{\alpha}\leq g(\mc,\nc|\rho)^{2(1-\beta)/\beta}{\,}\|\qp\|_{\beta}
\ , \label{eqresol}
\end{equation}
where $1/\alpha+1/\beta=2$, $1/2<\beta<1$ and the function
$g(\mc,\nc|\rho)$ is defined by
\begin{equation}
g(\mc,\nc|\rho)\triangleq\max\left\{(p_iq_j)^{-1/2}
{\,}|\tr(\mm_i\nm_j\rho)|:{\>} p_i\neq0,{\,} q_j\neq0 \right\}
\ . \label{gfdef}
\end{equation}
\end{Lem}

{\bf Proof} We first consider the case when both the resolutions
$\mc=\{\mm_i\}$ and $\nc=\{\nm_j\}$ are orthogonal. For those
values of labels $i$ and $j$ that satisfy $\tr(\mm_i\rho)\neq0$
and $\tr(\nm_j\rho)\neq0$, we put (generally non-Hermitian) operators
\begin{equation}
\omega_i =\|\mm_i\sqrt{\rho}\|_F^{-1}{\,}\mm_i\sqrt{\rho}
\ , \quad \theta_j =\|\nm_j\sqrt{\rho}\|_F^{-1}{\,}\nm_j\sqrt{\rho}
\ . \label{osdef}
\end{equation}
It is clear that $\|\omega_i\|_F=1$ and $\|\theta_j\|_F=1$. Since
the resolutions $\{\mm_i\}$ and $\{\nm_j\}$ are orthogonal,
we further have
$\langle\omega_i{\,},\omega_k\rangle_{\rm{hs}}=\delta_{ik}$ and
$\langle\theta_j{\,},\theta_k\rangle_{\rm{hs}}=\delta_{jk}$. The
matrix elements of transformation ${\mathsf{T}}$ are then defined
by
\begin{equation}
t_{ij}=\langle\omega_i{\,},\theta_j\rangle_{\rm{hs}}=
\|\mm_i\sqrt{\rho}\|_F^{-1}\|\nm_j\sqrt{\rho}\|_F^{-1}\langle\mm_i\sqrt{\rho}{\,},\nm_j\sqrt{\rho}\rangle_{\rm{hs}}
\ . \label{tijdefm}
\end{equation}
We now rewrite (\ref{lintrans}) as
$y_i(x)=\langle\omega_i{\,},\sigma\rangle_{\rm{hs}}$, where
$\sigma=\sum_jx_j\theta_j$ by definition. Due to $\|\sigma\|_F^{{\,}2}\equiv\sum_j|x_j|^2$ and
$\sigma=\sum_iy_i\omega_i+\varpi$, where
$\langle\omega_i{\,},\varpi\rangle_{\rm{hs}}=0$ for all $i$, we
have
$\bigl\|\sum_iy_i\omega_i\bigr\|_F^{{\,}2}\leq\|\sigma\|_F^{{\,}2}$
and the precondition of Lemma \ref{riesz} herewith. So we apply (\ref{suppos})
to the special values $y'_i=\|\mm_i\sqrt{\rho}\|_F$
and $x'_j=\|\nm_j\sqrt{\rho}\|_F$. By the completeness
relation,
$$
\pbi\sqrt{\rho}=\sum\nolimits_k y'_k\omega_k=\sum\nolimits_jx'_j\theta_j \ ,
$$
whence
$y'_i=\sum\nolimits_j\langle\omega_i{\,},\theta_j\rangle_{\rm{hs}}{\,}x'_j$,
i.e. the values $y'_i$ are related to $x'_j$ via the
transformation ${\mathsf{T}}$ too. Since the operators $\mm_i$ and
$\nm_j$ are projective, $p_i=|y'_i|^2$ and $q_j=|x'_j|^2$, whence
$\|\pp\|_{\alpha}=\|{\mathsf{y}}'\|_{a}^2$ and
$\|\qp\|_{\beta}=\|{\mathsf{x}}'\|_{b}^2$ with $\alpha=a/2$,
$\beta=b/2$. The statement of Lemma \ref{riesz} then results in
the inequality
$\|\pp\|_{\alpha}^{1/2}\leq{\max}|\langle\omega_i{\,},\theta_j\rangle_{\rm{hs}}|^{(1-\beta)/\beta}{\,}\|\qp\|_{\beta}^{1/2}$,
under the conditions $1/\alpha+1/\beta=2$ and $1/2<\beta<1$.
Noting that the maximum of modulus of (\ref{tijdefm}) is actually
$g(\mc,\nc|\rho)$, we resolve the case when the resolutions $\mc$
and $\nc$ are both orthogonal. To generalize (\ref{eqresol}) to
the case of arbitrary two resolutions, we will use the method
proposed in \cite{krishna} and further developed in
\cite{rast102}. In the extended space $\hh\oplus\kk$, the
resolutions $\mc$ and $\nc$ are realized as new resolutions
$\widetilde{\mc}$ and $\widetilde{\nc}$ respectively, and the
$\widetilde{\mc}$ is now orthogonal by Naimark's extension (see,
e.g., Sect. 5.1 in \cite{watrous1}). It can be made in such a way
that for any density matrix $\rho$ on $\hh$,
$\tr(\mm_i\rho)=\tr(\widetilde{\mm}_i\widetilde{\rho})$,
$\tr(\nm_j\rho)=\tr(\widetilde{\nm}_j\widetilde{\rho})$, and
$\tr(\mm_i\nm_j\rho)=\tr(\widetilde{\mm}_i\widetilde{\nm}_j\widetilde{\rho})$
(the $\widetilde{\rho}$ is built from $\rho$ by adding zero rows
and columns). Hence we have the same values of entropies
\cite{rast102} and
$g(\mc,\nc|\rho)=g(\widetilde{\mc},\widetilde{\nc}|\widetilde{\rho})$.
By a similar extension of $\widetilde{\nc}$, the question is quite
reduced to the above case of two orthogonal resolutions. $\square$

Using simple algebra (see the proof of proposition 3 in
\cite{rast102}), the inequality (\ref{eqresol}) can be rewritten
as
\begin{equation}
R_{\alpha}(\mc|\rho)+R_{\beta}(\nc|\rho)\geq -2\ln g(\mc,\nc|\rho)
\ . \label{renuim}
\end{equation}
For a pure state, this relation in terms the R\'{e}nyi entropies
coincides with the one deduced in the previous work
\cite{rast102}. With respect to the spectral decomposition
$\rho=\sum_{\lambda}
\lambda{\,}\bsi_{\lambda}\bsi_{\lambda}^{\dagger}$, we define
\begin{equation}
f(\mc,\nc|\rho)=\max\left\{\bigl(p_i^{(\lambda)}q_j^{(\lambda)}\bigr)^{-1/2}
|\langle\mm_i\bsi_{\lambda}{\,},\nm_j\bsi_{\lambda}\rangle|:{\>}p_i^{(\lambda)}\neq0,{\,}q_j^{(\lambda)}\neq0\right\}
\   , \label{ffdef}
\end{equation}
where the probabilities
$p_i^{(\lambda)}=\bigl\|\mm_i^{1/2}\bsi_{\lambda}\bigr\|_2^2$ and
$q_j^{(\lambda)}=\bigl\|\nm_j^{1/2}\bsi_{\lambda}\bigr\|_2^2$. The
above inequality with $f(\mc,\nc|\rho)$ instead of
$g(\mc,\nc|\rho)$ was obtained in \cite{rast102}. For an impure
state, we have $g(\mc,\nc|\rho)<f(\mc,\nc|\rho)$ and a stronger
bound in (\ref{renuim}). For canonically conjugate variables, the uncertainty
relations in terms of R\'{e}nyi's entropies were given in \cite{birula3}.

Let us proceed to a relation with the
Tsallis entropies. The sum $H_{\alpha}(p_i)+H_{\beta}(q_j)$ cannot
be arbitrarily small because of the constraint (\ref{eqresol}). We
rewrite this sum as
\begin{equation}
H_{\alpha}(p_i)+H_{\beta}(q_j)=\phi(\xi,\zeta)=\frac{\xi-1}{1-\alpha}+\frac{\zeta-1}{1-\beta}
\label{sumhh}
\end{equation}
in terms of the variables
$\xi=\sum_ip_i^{\alpha}=\|\pp\|_{\alpha}^{\alpha}$,
$\zeta=\sum_jq_j^{\beta}=\|\qp\|_{\beta}^{\beta}$ and the function
$\phi(\xi,\zeta)$. Assuming $\alpha>1>\beta$, we obviously have $\xi\leq1$
and $\zeta\geq1$. Adding (\ref{eqresol}) in the form
$\zeta\geq\gamma{\,}\xi^{\beta/\alpha}$, where
$\gamma=g(\mc,\nc|\rho)^{-2(1-\beta)}$, we have arrived at a task
of minimizing $\phi(\xi,\zeta)$ under the above conditions. It is
important here that $g(\mc,\nc|\rho)\leq1$ and, therefore,
$\gamma\geq1$. Indeed, the Cauchy-Schwarz inequality for the
Hilbert-Schmidt inner product directly shows that the modulus of
(\ref{tijdefm}) does not exceed one. The task is solved in
Appendix \ref{app1}, and the sum $H_{\alpha}(p_i)+H_{\beta}(q_j)$
cannot be less than
$$
\phi(\xi_0,1)=\frac{\gamma^{-\alpha/\beta}-1}{1-\alpha}=\frac{g(\mc,\nc|\rho)^{2\alpha(1-\beta)/\beta}-1}{1-\alpha}=
\frac{g(\mc,\nc|\rho)^{2(\alpha-1)}-1}{1-\alpha}=\ln_{\alpha}\Bigl(g(\mc,\nc|\rho)^{-2}\Bigr) \ ,
$$
where we used $(1-\beta)/\beta=(\alpha-1)/\alpha$ due to
$1/\alpha+1/\beta=2$. By $g(\mc,\nc|\rho)=g(\nc,\mc|\rho)$, we claim the following.

\begin{Thm}\label{tsasol}
For any two resolutions $\mc=\{\mm_i\}$ and $\nc=\{\nm_j\}$ of the
identity and given density operator $\rho$, there holds
\begin{equation}
H_{\alpha}(\mc|\rho)+H_{\beta}(\nc|\rho) \geq \ln_{\mu}\Bigl(g(\mc,\nc|\rho)^{-2}\Bigr)
\ , \label{eqtsasol}
\end{equation}
where $1/\alpha+1/\beta=2$ and $\mu=\max\{\alpha,\beta\}$.
\end{Thm}

The inequality (\ref{eqtsasol}) gives the uncertainty relation in
terms of the Tsallis entropies for two generalized measurements.
In spirit and origin, it is like to the relation via the R\'{e}nyi
entropies (\ref{renuim}). Note that relations in terms of the
Tsallis entropies have been considered for particular cases of the
position and momentum \cite{raja,wlwl09} and the spin-1/2
components \cite{majer,shpyrko}. It is of some interest to get a
state-independent version of (\ref{eqtsasol}). Since the function
$\ln_{\mu}x$ is increasing for $\mu>0$,
$g(\mc,\nc|\rho)\leq f(\mc,\nc|\rho)$ and \cite{rast102}
\begin{equation}
f(\mc,\nc|\bsi)\leq\max\Bigl\{\bigl\|\mm_i^{1/2}\nm_j^{1/2}\bigr\|_{\infty}:
{\ }\mm_i\in\mc,{\,}\nm_j\in\nc\Bigr\}\triangleq\bar{f}(\mc,\nc)
\ , \label{deffbar}
\end{equation}
we have the state-independent bound
\begin{equation}
H_{\alpha}(\mc|\rho)+H_{\beta}(\nc|\rho) \geq \ln_{\mu}\Bigl(\bar{f}(\mc,\nc)^{-2}\Bigr)
\ . \label{eqtsind}
\end{equation}
Note that $\bar{f}(\mc,\nc)\leq1$ is provided by
$\bigl\|\mm_i^{1/2}\nm_j^{1/2}\bigr\|_{\infty}^2\leq\|\mm_i\|_{\infty}{\,}\|\nm_j\|_{\infty}\leq1$.
Here the inequality on the left is a Cauchy-Schwarz inequality for
ordinary matrix products and the spectral norm (see, e.g., the result (4.50)
in \cite{zhan02}) and the inequality on the right follows from the
completeness relation. The inequality $\bar{f}(\mc,\nc)\leq1$ is
saturated if and only if for some two elements $\mm_0$, $\nm_0$
and nonzero $\bhi_0\in\hh$ there hold $\mm_0\bhi_0=\bhi_0$ and
$\nm_0\bhi_0=\bhi_0$ simultaneously. In other words,
$\bar{f}(\mc,\nc)=1$ implies that there exist those operators
$\mm_0\in\mc$ and $\nm_0\in\nc$ that act as commuting projectors
in a nonempty subspace $\kk_0\subset\hh$. Otherwise, the entropic
bound in the relation (\ref{eqtsind}) is nontrivial. So, we have
extended the relation, conjectured in \cite{kraus87} and later
proved in \cite{maass}, to the Tsallis entropies and general
quantum measurements. Let us consider two concrete examples of
specific interest.

\begin{Exam}\label{fdisc}
The first example is a pair of complementary observables in a
$d$-level system (for the R\'{e}nyi formulation, see
\cite{birula3}). Let complex amplitudes $\tilde{c}_k$ and $c_l$ be
connected by the discrete Fourier transform
\begin{equation}
\tilde{c}_k = \frac{1}{\sqrt{d}}{\>}\sum\nolimits_{l=1}^d e^{2\pi{\ri} k{\,}l/d}{\>} c_l
\ , \label{discf}
\end{equation}
and the corresponding probability distributions
$p_k=|\tilde{c}_k|^2$ and $q_l=|c_l|^2$. The transformation
(\ref{discf}) is the ''canonical'' example that leads to
complementary observables \cite{kraus87}. It follows from
$\|\mathsf{\tilde{c}}\|_2=\|\cp\|_2$ and (\ref{suppos}) that
\begin{equation}
\|\cp\|_{a}\leq  \left(\frac{1}{\sqrt{d}}\right)^{(2-b)/b} \|\mathsf{\tilde{c}}\|_{b}
\ , {\ }\qquad
\|\mathsf{\tilde{c}}\|_{a}\leq \left(\frac{1}{\sqrt{d}}\right)^{(2-b)/b} \|\cp\|_{b}
\ , \label{cacb}
\end{equation}
where $1/a+1/b=1$ and $1<b<2$. Squaring, we get 
$\|\qp\|_{\alpha}\leq(1/d)^{(1-\beta)/\beta}\|\pp\|_{\beta}{\,}$,
$\|\pp\|_{\alpha}\leq(1/d)^{(1-\beta)/\beta}\|\qp\|_{\beta}$
under the conditions on $\alpha$ and $\beta$ from Lemma \ref{resol}. In
its symmetric form, the uncertainty relation in terms of the
Tsallis entropies is written as
$H_{\alpha}(p_k)+H_{\beta}(q_l)\geq\ln_{\mu}d{\,}$, where
$1/\alpha+1/\beta=2$ and $\mu=\max\{\alpha,\beta\}$.
\end{Exam}

\begin{Exam}\label{angl}
The angle $\phi$ and the angular momentum $\jm_z$ can similarly be
treated. Taking one and the same size $\delta\varphi$ for all
the angular bins (i.e. the ratio $2\pi/\delta\varphi$ is a
strictly positive integer), we introduce probabilities
\begin{equation}
p_k=\int_{k\delta\varphi}^{(k+1)\delta\varphi} d\varphi {\,}|\Psi(\varphi)|^2
\ , {\ }\qquad
q_l=|c_l|^2
\ , \label{anprob}
\end{equation}
where the coefficients $c_l$'s are related to the expansion
$\Psi(\varphi)=(2\pi)^{-1/2}\sum_{l=-\infty}^{+\infty}c_l\exp(\ri{\,}l{\,}\varphi)$,
with respect to the eigenstates of the $\jm_z$. Using theorem 192 of the book 
\cite{hardy} for integral means and assuming $\beta<1<\alpha$, we have
\begin{equation}
\frac{1}{\delta\varphi}\int_{k\delta\varphi}^{(k+1)\delta\varphi} d\varphi {\,}|\Psi(\varphi)|^{2\beta}
\leq
\left(\frac{1}{\delta\varphi}\int_{k\delta\varphi}^{(k+1)\delta\varphi} d\varphi {\,}|\Psi(\varphi)|^{2}\right)^{\beta}
\label{conal}
\end{equation}
and the inversed inequality with $\alpha$ instead of $\beta$. Summing these inequalities with respect to $k$ and then raising them to the powers $1/\beta$ and $1/\alpha$ respectively, we finally write 
\begin{equation}
\|\Psi\|_b^2\leq\delta\varphi^{(1-\beta)/\beta}\|\pp\|_{\beta}
\ , {\ }\qquad
\delta\varphi^{(1-\alpha)/\alpha}\|\pp\|_{\alpha}\leq\|\Psi\|_a^2	
\ , \label{cor192} 
\end{equation}
where the norm
$\|\Psi\|_b=\left(\int_{0}^{2\pi}d\varphi {\,}|\Psi(\varphi)|^b\right)^{1/b}$ and
$b=2\beta$, $a=2\alpha$. Combining the relations (\ref{cor192}) with the Young-Hausdorff
inequalities (see, e.g., section 8.17 in \cite{hardy}), which are viewed in our notation as
\begin{equation}
\|\cp\|_{a}\leq\left(\frac{1}{\sqrt{2\pi}}\right)^{(2-b)/b} \|\Psi\|_b
\ , {\ }\qquad
\|\Psi\|_a\leq\left(\frac{1}{\sqrt{2\pi}}\right)^{(2-b)/b} \|\cp\|_{b}
\ , \label{yona}
\end{equation}
we get
$\|\qp\|_{\alpha}\leq\bigl(\delta\varphi/2\pi\bigr)^{(1-\beta)/\beta}\|\pp\|_{\beta}{\,}$,
$\|\pp\|_{\alpha}\leq\bigl(\delta\varphi/2\pi\bigr)^{(1-\beta)/\beta}\|\qp\|_{\beta}{\,}$.
So, there holds
$H_{\alpha}(\varphi)+H_{\beta}(\jm_z)\geq\ln_{\mu}\bigl(2\pi/\delta\varphi\bigr)$,
where $1/\alpha+1/\beta=2$ and $\mu=\max\{\alpha,\beta\}$. When
$\mu\to1$, this new inequality in terms of Tsallis' entropies
coincides with the relation in terms of R\'{e}nyi's entropies
deduced in \cite{birula3}.
\end{Exam}

Finally, we consider the case of extremal unravelings within the
general formulation (\ref{eqtsasol}). For the extremal unravelings
$\ac_{\rho}^{(ex)}$ and $\bc_{\rho}^{(ex)}$ of the super-operators
$\$_A$ and $\$_B$, we obtain the entropic uncertainty relation
\begin{equation}
H_{\alpha}(\ac_{\rho}^{(ex)}|\rho)+H_{\beta}(\bc_{\rho}^{(ex)}|\rho)
\geq \ln_{\mu}\Bigl(g\bigl(\ac_{\rho}^{(ex)},\bc_{\rho}^{(ex)}|\rho\bigr)^{-2}\Bigr)
\ , \label{enunun}
\end{equation}
where the $g(\ac_{\rho}^{(ex)},\bc_{\rho}^{(ex)}|\rho)$ is put by
(\ref{gfdef}) with $\mm_i=\am_i^{\dagger}\am_i$,
$\nm_j=\bn_j^{\dagger}\bn_j$ in terms of Kraus operators
$\am_i\in\ac_{\rho}^{(ex)}$, $\bn_j\in\bc_{\rho}^{(ex)}$. We can
also write the uncertainty relation for unravelings extremal with
respect to the R\'{e}nyi entropies. Using (\ref{renuim}), for
$\alpha>1$ we obtain the relation
\begin{equation}
R_{\alpha}\bigl({\ac'}_{\alpha,\rho}^{(ex)}|\rho\bigr)+R_{\beta}\bigl(\bc_{\rho}^{(ex)}|\rho\bigr) \geq
-2 \ln g\bigl({\ac'}_{\alpha,\rho}^{(ex)},\bc_{\rho}^{(ex)}|\rho\bigr)
\ , \label{renunn}
\end{equation}
where ${\ac'}_{\alpha,\rho}^{(ex)}$ denotes the unraveling of
$\$_A$ extremal for the R\'{e}nyi entropy of order $\alpha>1$.
This unraveling differs from the one given by (\ref{excalc}) and
also depends on the parameter $\alpha$ in general. A search of
explicit analytic expression for ${\ac'}_{\alpha,\rho}^{(ex)}$
seems to be complicated, because concavity (convexity) things
cannot be used here. Nevertheless, the entropic relation in terms of
the R\'{e}nyi entropies holds, as a mathematical inequality at least.

\section{Conclusion}\label{sec4}

We have considered the unraveling (i.e. the concrete set of Kraus
operators) of a super-operator that is extremal with respect to
all the Tsallis entropies of a positive order and the R\'{e}nyi
entropies of an order $0<\alpha<1$. This general result is
formally posed in Theorem \ref{extr}. If one of unravelings is
given explicitly, then this minimizing unraveling is easily
calculated by diagonalizing a certain Hermitian matrix. The known
relation between the Shannon entropy of an ensemble of pure states
and the von Neumann entropy of the risen density operator is
extended to both the Tsallis and R\'{e}nyi entropies of the
mentioned orders. The case of Tsallis entropy allows further
extension to a mixture of density operators (see the bounds
(\ref{concmix})). Due to Riesz's theorem, there exists an
inequality between certain functions of the probability
distributions generated by two resolutions of the identity (see
Lemma \ref{resol}). This inequality gives an origin for the
uncertainty relations, given by (\ref{renuim}) for the R\'{e}nyi
entropies and by Theorem \ref{tsasol} for the Tsallis entropies.
The general formulation (\ref{eqtsasol}) is a new result in the
topic. It has been illustrated within the two interesting cases
(see Examples \ref{fdisc} and \ref{angl}). In the formulae
(\ref{enunun}) and (\ref{renunn}), both the entropic uncertainty
relations are naturally recast for the extremal unravelings of two
given trace-preserving super-operators.

\appendix

\section{A minimum of the function}\label{app1}

To obtain a lower bound on the sum of Tsalis entropies, we find
the minimal value of the function (\ref{sumhh}) in the domain
$D$ such that $0\leq\xi\leq1$, $1\leq\zeta<\infty$,
$\zeta\geq\gamma{\,}\xi^{\beta/\alpha}$. When $\gamma>1$, the
curve $\zeta=\gamma{\,}\xi^{\beta/\alpha}$ cuts off the down right
corner of the rectangle
$\{(\xi,\zeta):{\>}0\leq\xi\leq1,{\,}1\leq\zeta<\infty\}$ and
herewith the point $(1,1)$ in which $\phi=0$ (see Fig. 1). In the
interior of $D$, we have
\begin{equation}
\frac{\partial\phi}{\partial\xi}=\frac{1}{1-\alpha}<0  \ , \qquad
\frac{\partial\phi}{\partial\zeta}=\frac{1}{1-\beta}>0  \ ,
\label{partder}
\end{equation}
due to $\alpha>1>\beta$. So the minimum is reached on the boundary
of the domain $D$. Using (\ref{partder}), the task
is merely reduced to minimizing $\phi(\xi,\zeta)$ on segment
of the curve $C:{\>}\zeta=\gamma{\,}\xi^{\beta/\alpha}$ between the
point $(\xi_0,1)$, where $\xi_0=\gamma^{-\alpha/\beta}$, and the
point $(1,\gamma)$. Substituting
$\zeta=(\xi/\xi_0)^{\beta/\alpha}$ in (\ref{sumhh}) and
differentiating with respect to $\xi$, we obtain
\begin{equation}
\frac{1}{1-\alpha}+\frac{\beta}{\alpha(1-\beta)\xi}\left(\frac{\xi}{\xi_0}\right)^{\beta/\alpha}=
\frac{1}{\alpha-1}\left(\frac{1}{\xi}\left(\frac{\xi}{\xi_0}\right)^{\beta/\alpha}-{\,}1 \right)
\ , \label{deriv}
\end{equation}
where we used $\beta/(1-\beta)=\alpha/(\alpha-1)$ because of
$1/\alpha+1/\beta=2$. When $\xi_0<1$, the quantity (\ref{deriv})
is strictly positive for $\xi_0\leq\xi\leq1$ and the minimal value
is $\phi(\xi_0,1)=(\xi_0-1)/(1-\alpha)$ too.

\begin{figure*} 
\includegraphics[width=4.5cm]{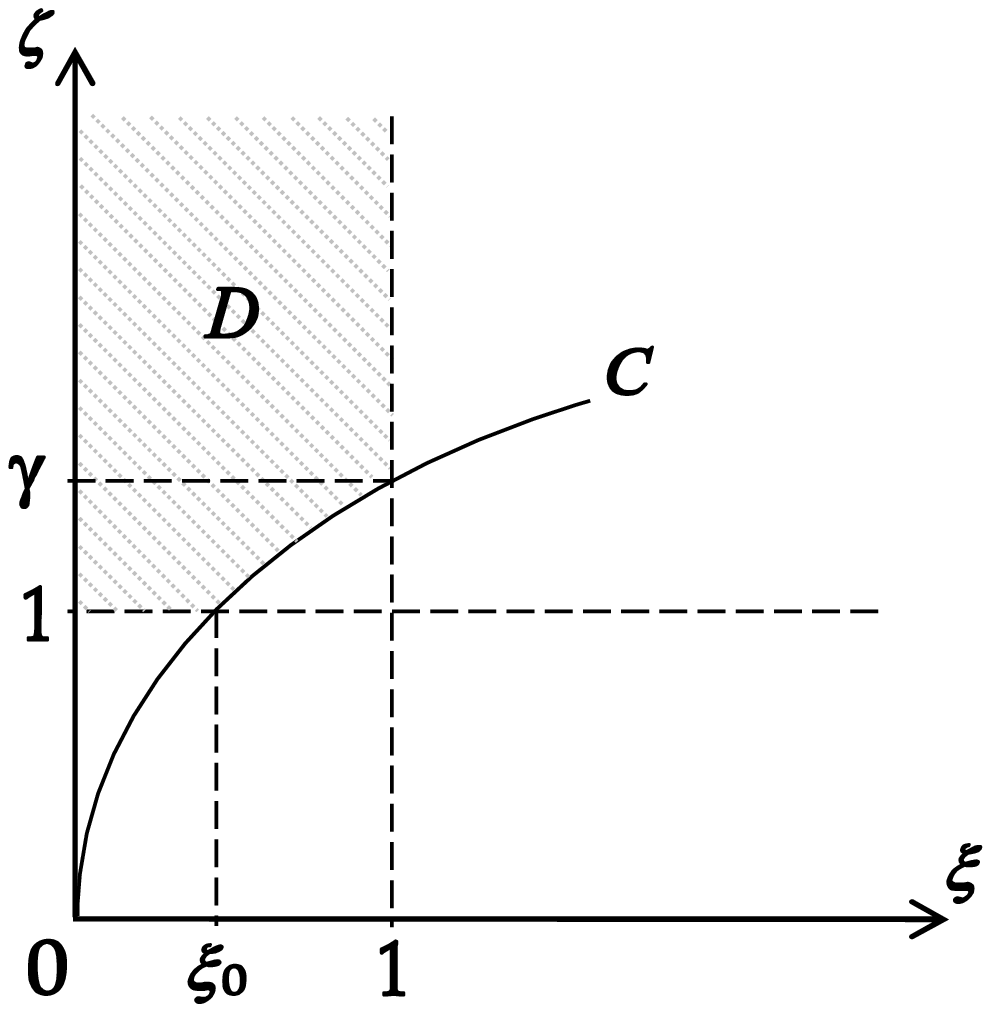}
\caption{The domain $D$ in which we find the minimum of the function $\phi(\xi,\zeta)$ defined by (\ref{sumhh}).}
\end{figure*}


\begin{thebibliography}{59}

\bibitem{heisenberg}%---------------------------------------------
Heisenberg W 1927 {\it Z. Phys.} {\bf 43} 172

\bibitem{hall99}%------------------------------------------------
Hall M J W 1999 {\it Phys. Rev.} A {\bf 59} 2602

\bibitem{lahti}%-----------------------------------------------
Busch P, Heinonen T and Lahti P J 2007 {\it Phys. Rep.} {\bf 452} 155

\bibitem{robert}%--------------------------------------------------
Robertson H P 1929 {\it Phys. Rev.} {\bf 34} 163

\bibitem{lynch}%--------------------------------------------------
Lynch R 1995 {\it Phys. Rep.} {\bf 256} 357

\bibitem{barnett}%--------------------------------------------------
Barnett S M and Pegg D T 1989 {\it J. Mod. Optics} {\bf 36} 7

\bibitem{hirs}%-------------------------------------------------
Hirschman I I 1957 {\it Am. J. Math.} {\bf 79} 152

\bibitem{maass}%----------------------------------------------------
Maassen H and Uffink J B M 1988 {\it Phys. Rev. Lett.} {\bf 60} 1103

\bibitem{ww10}%-------------------------------------------------
Wehner S and Winter A 2010 {\it New J. Phys.} {\bf 12} 025009

\bibitem{mamojka}%--------------------------------------------------
Mamojka B 1974 {\it Int. J. Theor. Phys.} {\bf 11} 73

\bibitem{deutsch}%---------------------------------------------------
Deutsch D 1983 {\it Phys. Rev. Lett.} {\bf 50} 631

\bibitem{spindel}%-------------------------------------------------
Massar S and Spindel P 2008 {\it Phys. Rev. Lett.} {\bf 100} 190401

\bibitem{lanz}%-------------------------------------------------
Urizar-Lanz I and T\'{o}th G 2010 {\it Phys. Rev. A} {\bf 81} 052108

\bibitem{ivan92}%---------------------------------------------------
Ivanovic I D 1992 {\it J. Phys. A: Math. Gen.} {\bf 25} L363

\bibitem{sanchez93}%---------------------------------------------------
Sanchez J 1993 {\it Phys. Lett. A} {\bf 173} 233

\bibitem{hall08}%-------------------------------------------------
Hall M J W 2008 {\it J. Phys. A: Math. Theor.} {\bf 41} 255301

\bibitem{manko09}%----------------------------------------------
Man'ko M A 2009 {\it J. Russ. Laser Res.} {\bf 30} 514

\bibitem{kraus}%-------------------------------------------------
Kraus K 1983 {\it States, Effects, and Operations: Fundamental Notions of Quantum Theory} ({\it Lecture Notes
in Physics} vol 190) (Berlin: Springer)

\bibitem{ilichev03}%------------------------------------------------
Il'ichev L V 2003 {\it JETP} {\bf 96} 982

\bibitem{tsbibl}%------------------------------------------------
http://tsallis.cat.cbpf.br/biblio.htm

\bibitem{aczel}%-------------------------------------------------
Acz\'{e}l J and Dar\'{o}czy Z 1975 {\it On Measures of Information and their Characterizations} (New York: Academic Press)

\bibitem{tsallis}%-------------------------------------------------
Tsallis C 1988 {\it J. Stat. Phys.} {\bf 32} 479

\bibitem{havrda}%-------------------------------------------------
Havrda J and Charv\'{a}t F 1967 {\it Kybernetika} {\bf 3} 30

\bibitem{renyi}%-------------------------------------------------
R\'{e}nyi A 1961 {\it On measures of entropy and information},
Proceedings of the 4th Berkeley Symposium on Mathematical Statistics
and Probability, pp. 547-561 (University of California Press:
Berkeley--Los Angeles)

\bibitem{zycz}%----------------------------------------------------
\.{Z}yczkowski K 2003 {\it Open Sys. Inf. Dyn.} {\bf 10} 297

\bibitem{watrous1}%------------------------------------------------
Watrous J 2008 {\it CS 798: Theory of Quantum Information} (University of Waterloo:
\\ http://www.cs.uwaterloo.ca/$\sim$watrous/quant-info/lecture-notes/all-lectures.pdf)

\bibitem{holevo}%----------------------------------------------------
Holevo A S 1982 {\it Probabilistic and Statistical Aspects of Quantum Theory}
({\it North-Holland Series in Statistics and Probability} vol 1)
(Amsterdam: North-Holland)

\bibitem{preskill}%------------------------------------------------
Preskill J 1998 {\it Lecture Notes for Physics 229: Quantum Computation and Information} (California
Institute of Technology: \\ http://www.theory.caltech.edu/people/preskill/ph229/)

\bibitem{carm}%------------------------------------------------
Carmichael H J 1993 {\it An Open Systems Approach to Quantum Optics} ({\it Lecture Notes
in Physics} vol m18) (Berlin: Springer)

\bibitem{hardy}%----------------------------------------------------
Hardy G H, Littlewood J E and Polya G 1934 {\it Inequalities}
(London: Cambridge University Press)

\bibitem{breslin}%----------------------------------------------------
Breslin J K and Milburn G J 1997 {\it J. Mod. Optics} {\bf 44} 2469

\bibitem{hugston}%--------------------------------------------------
Hughston L P, Jozsa R and Wootters W K 1993 {\it Phys. Lett. A} {\bf 183} 14

\bibitem{wehrl}%--------------------------------------------------
Wehrl A 1978 {\it Rev. Mod. Phys.} {\bf 50} 221

\bibitem{rast10fn}%----------------------------------------------------
Rastegin A E 2010 Fano type quantum inequalities in terms of $q$-entropies {\it E-print} arXiv:1010.1811 [quant-ph]

\bibitem{rast102}%----------------------------------------------------
Rastegin A E 2010 {\it J. Phys. A: Math. Theor.} {\bf 43} 155302

\bibitem{krishna}%-------------------------------------------------
Krishna M and Parthasarathy K R 2002 {\it Sankhya, Ser. A} {\bf 64} 842

\bibitem{birula3}%------------------------------------------------
Bialynicki-Birula I 2006 {\it Phys. Rev.} A {\bf 74} 052101

\bibitem{raja}%-------------------------------------------------
Rajagopal A K 1995 {\it Phys. Lett.} A {\bf 205} 32

\bibitem{wlwl09}%-----------------------------------------------
Wilk G and W{\l}odarczyk Z 2009 {\it Phys. Rev.} A {\bf 79} 062108

\bibitem{majer}%-------------------------------------------------
Majernik V and Majernikova E 2001 {\it Rep. Math. Phys.} {\bf 47} 381

\bibitem{shpyrko}%-------------------------------------------------
Majernik V,  Majernikova E and Shpyrko S 2003 {\it Cent. Eur. J. Phys.} {\bf 3} 393

\bibitem{zhan02}%----------------------------------------------
Zhan X 2002 {\it Matrix Inequalities} (Berlin: Springer)

\bibitem{kraus87}%----------------------------------------------
Kraus K 1987 {\it Phys. Rev.} D {\bf 35} 3070

\end{thebibliography}
\end{document}